\documentclass[9pt,note]{IEEEtran}
\usepackage{amsmath,amssymb,amsfonts,theorem}
\usepackage{graphicx,epsfig,psfrag}
\usepackage{subfigure}
\usepackage{wasysym}
\usepackage{mathtools}

\usepackage{float}
\usepackage{cite}
\usepackage[prependcaption,colorinlistoftodos]{todonotes}

\usepackage{tikz}
\usepackage{tkz-berge}
\usetikzlibrary{arrows,%
                petri,%
                topaths}%
\usetikzlibrary{decorations.markings}         
\tikzstyle{vertex}=[circle, shading = ball, ball color = white!100!white, minimum size = 10pt, inner sep = 1pt, draw, inner sep=0pt]  
\newcommand{\vertex}{\node[vertex]}           

\newcommand{\lmin}{\ell_{\mathrm{min}}}

\let\leq\leqslant

\newcommand{\calD}{\ensuremath{\mathcal{D}}}

\newcommand{\calQ}{\ensuremath{\mathcal{Q}}}

\newcommand{\hatX}{\ensuremath{\hat{X}}}

\newcommand{\bmat}{\begin{matrix}}
\newcommand{\emat}{\end{matrix}}
\newcommand{\bbm}{\begin{bmatrix}}
\newcommand{\ebm}{\end{bmatrix}}
\newcommand{\bpm}{\begin{pmatrix}}
\newcommand{\epm}{\end{pmatrix}}
\newcommand{\bse}{\begin{subequations}}
\newcommand{\ese}{\end{subequations}}
\newcommand{\beq}{\begin{equation}}
\newcommand{\eeq}{\end{equation}}
\newcommand{\ben}{\begin{enumerate}}
\newcommand{\een}{\end{enumerate}}
\newcommand{\beni}{\renewcommand{\labelenumi}{\roman{enumi}.}
\renewcommand{\theenumi}{\roman{enumi}}\begin{enumerate}}
\newcommand{\eeni}{\end{enumerate}\renewcommand{\labelenumi}{\arabic{enumi}.}
\renewcommand{\theenumi}{\arabic{enumi}}}
\newcommand{\bena}{\renewcommand{\labelenumi}{\alpha{enumi}.}
\renewcommand{\theenumi}{\alpha{enumi}}\begin{enumerate}}
\newcommand{\eena}{\end{enumerate}\renewcommand{\labelenumi}{\arabic{enumi}.}
\renewcommand{\theenumi}{\arabic{enumi}}}
\newcommand{\bit}{\begin{itemize}}
\newcommand{\eit}{\end{itemize}}
\newcommand{\bthe}{\begin{theorem}}
\newcommand{\ethe}{\end{theorem}}
\newcommand{\blem}{\begin{lemma}}
\newcommand{\elem}{\end{lemma}}
\newcommand{\bprop}{\begin{proposition}}
\newcommand{\eprop}{\end{proposition}}
\newcommand{\bex}{\begin{example}}
\newcommand{\eex}{\end{example}}
\newcommand{\bas}{\begin{assumption}}
\newcommand{\eas}{\end{assumption}}
\newcommand{\bre}{\begin{remark}}
\newcommand{\ere}{\end{remark}}
\newcommand{\bcor}{\begin{corollary}}
\newcommand{\ecor}{\end{corollary}}
\newcommand{\bdfn}{\begin{definition}}
\newcommand{\edfn}{\end{definition}}

\newcommand{\ones}{\ensuremath{1\!\!1}}

\newcommand{\pset}[1]{\ensuremath{\{#1\}}}
\newcommand{\nset}[1]{\ensuremath{\{1,2,\ldots,#1\}}}

\newcommand{\abs}[1]{\ensuremath{| #1 |}}

\newcommand{\R}{\ensuremath{\mathbb R}}

\newcommand{\BP}{\noindent{\bf Proof. }}
\newcommand{\EP}{\hspace*{\fill} $\blacksquare$\bigskip\noindent}

{\theorembodyfont{\itshape}\newtheorem{theorem}{Theorem}[section]}
{\theorembodyfont{\itshape}\newtheorem{proposition}[theorem]{Proposition}}
{\theorembodyfont{\itshape}\newtheorem{lemma}[theorem]{Lemma}}
{\theorembodyfont{\itshape}\newtheorem{corollary}[theorem]{Corollary}}
{\theorembodyfont{\upshape}\newtheorem{definition}[theorem]{Definition}}
{\theorembodyfont{\itshape}}
{\theorembodyfont{\upshape}\newtheorem{remark}[theorem]{Remark}}
{\theorembodyfont{\upshape}\newtheorem{example}[theorem]{Example}}

\newcommand{\Qyek}{\calQ_s(G_1)}
\newcommand{\Qdo}{\calQ_{ss}(G_1)}
\begin{document}
\date{}
\title{Zero forcing sets and controllability of dynamical systems defined on graphs}

%
\author{
Nima Monshizadeh\authorrefmark{1} \thanks{$^*\;$ Johann Bernoulli Institute for Mathematics and Computer Science, University of Groningen, P. O Box 800, 9700 AV Groningen, The Netherlands, Phone:+31-50-3633999, Email: {\tt\small n.monshizadeh@rug.nl}, and {\tt\small{m.k.camlibel@rug.nl}.}
}\and Shuo Zhang\authorrefmark{2}\thanks{$^\dagger\;$Research Institute of Industrial Technology and Management, University of Groningen, 9747AG, Groningen, the Netherlands, Phone: +310503634771, Email: {\tt\small shuo.zhang@rug.nl}} \and M. Kanat Camlibel\authorrefmark{1}$^,$\authorrefmark{3} \thanks{$^\ddagger\;$Dept. of Electronics Eng., Dogus University, Kadikoy, Istanbul, Turkey.}}

\maketitle{}

\begin{abstract}
In this paper, controllability of systems defined on graphs is discussed.
We consider the problem of controllability of the network for a family of matrices carrying the structure of an underlying directed graph.
A one-to-one correspondence between the set of leaders rendering the network controllable and zero forcing sets is established.
To illustrate the proposed results, special cases including path, cycle, and complete graphs are discussed.
Moreover,  as shown for graphs with a tree structure, the proposed results of the present paper together with the existing results on the zero forcing sets lead to a minimal leader selection scheme in particular cases.
\end{abstract}

\section{Introduction}

The study of networks of dynamical systems became one of the most popular themes within systems and control theory in the last two decades. Roughly speaking, networks of dynamical systems can be seen as dynamical systems that inherit certain structural properties from the topology of a graph that captures the network structure. Across many scientific disciplines, one encounters such systems in a variety of applications. Typical examples include biological, chemical, social, power grid, and robotic networks (see e.g. \cite[Ch. 1]{MMME10}). The research on numerous aspects of these kind of systems have already resulted in a vast literature that still keeps growing.

One line of research in this fast growing literature is devoted to the controllability analysis of linear input/state systems of the form
$$
\dot{x}=Xx+Uu
$$
where $x\in\R^n$ is the state and $u\in\R^m$ is the input with the distinguishing
feature that the matrix $X$ is associated with a given graph and the matrix $U$ encodes the vertices (often called leaders) through which external inputs are applied.

Up to our knowledge, \cite{HGT04} is the first paper which addressed controllability problem within this framework when $X$ is the Laplacian matrix of an undirected graph. This early paper was followed by a number of papers dealing with different aspects of controllability when $X$ is the Laplacian matrix (see e.g.\cite{ARMJMMME09}, \cite{MESMMCMKCAB12}, \cite{SZMKCMC11}) and when $X$ is the adjacency matrix (see e.g. \cite{CDG10}). On the one hand, controllability was investigated from a graph topology perspective in \cite{ARMJMMME09}, \cite{MESMMCMKCAB12}, \cite{SMMEAB10}, \cite{MKCSZMC12}, \cite{SZMKCMC11}, \cite{AYYWAME12}, \cite{CDG10} which established necessary/sufficient conditions for controllability as well as lower and/or upper bounds on the controllable subspace. These conditions are based on graph theoretical tools such as graph symmetry \cite{ARMJMMME09}, (almost) equitable partitions \cite{ARMJMMME09}, \cite{MESMMCMKCAB12}, \cite{SMMEAB10}, \cite{SZMKCMC11}, walks of a graph \cite{CDG10}, distance partitions \cite{SZMKCMC11}, or pseudo monotonically increasing sequences \cite{AYYWAME12}.
On the other hand, the minimum number of leaders that render the system controllable, with $X$ being the Laplacian matrix of a simple undirected graph, was explored for several classes of graphs such as path graphs \cite{ARMJMMME09}, \cite{GPGN12}, cycle graphs \cite{GPGN12}, \cite{SZMKCMC11}, complete graphs \cite{ARMJMMME09}, \cite{SZMKCMC11}, and circulant graphs \cite{MNMMta} which all provide also a leader selection procedure.

Another thread in the study of controllability of systems defined by a graph was centered around structural controllability. Structural controllability deals with a family of pairs $(X,U)$ rather than a particular instance and asks whether the family contains a controllable pair (weak structural controllability \cite{YYLJJSALB11}) or all members of the family are controllable (strong structural controllability \cite{ACMMs}). In the latter case, the authors of \cite{ACMMs} have established necessary and sufficient conditions for strong structural controllability in terms of constrained matchings over the bipartite graph representation of the network. For a more general look at control properties of structured linear systems, see e.g. \cite{JMDCCJW:03}.

In this paper, we deal with a family of $X$ matrices carrying the structure of a directed graph $G$. This family is called the qualitative class of $G$, and  we investigate the controllability of the network with respect to this qualitative class, under a fixed set of vertices (leaders). Note that essentially this is the same as studying strong structural controllability, but we carry out controllability analysis through the notion of zero forcing sets, similar to \cite{DBDDLHSSMYta}, rather than through the constrained matching which has been treated in \cite{ACMMs}.

The notions of zero forcing sets and zero forcing number have an intimate relationship with minimum rank problems of patterned matrices, and have been well studied in the literature (see e.g. \cite{FBWBSMFHTHLHBSPDHH10} and \cite{LH10}). Moreover, in these papers and the references therein, lower/upper bounds for the zero forcing number has been provided, and also the exact value has been obtained for some special classes of graphs, either directly or in terms of some graph parameters such as path cover number. Note that computing the zero forcing number as well as finding a minimal zero forcing set for a general loop directed graph is an NP-hard problem (see \cite[Thm. 2.6]{MTJCDs}).

Recently, zero forcing sets in one form or another have been utilized for controllability analysis of quantum systems as well as linear systems (see e.g. \cite{DBDDLHSSMYta}, \cite{DBVG:07}, \cite{DBSBCBVG09}). In particular, for the case where the underlying communication graph is undirected and all off-diagonal elements of $X$ have the same sign, a sufficient condition for network controllability and in terms of zero forcing sets has been provided in \cite{DBDDLHSSMYta}.

In this paper, for the case where the underlying graph is directed, we establish a one-to-one correspondence between the set of leaders rendering the network controllable and zero forcing sets.
Consequently, we obtain that the minimum number of leaders required to render the network controllable, with respect to the whole qualitative class, is indeed equal to the zero forcing number of the underlying graph. Note that in some applications extra assumptions and constraints such as symmetry may be present on the entries of the matrix $X$. Hence, in these cases, one may be interested in some subsets of the qualitative class of $G$ rather than the whole class.  This will be addressed through the notion of sufficiently rich subclasses, and we explore how the results established in this paper boils down or can be applied to certain qualitative subclasses.
Then, we study the controllability problem for some special classes of graphs, namely path, cycle, and complete graphs.
In addition, we establish a connection between the existing results on the minimum number of leaders in these cases where the matrix $X$ is the Laplacian matrix, and the results proposed in this manuscript.

An advantage of the proposed results of this paper is that one can deduce conclusions on the minimum number of leaders for controllability as well as how to choose such leaders in particular cases, by utilizing the existing results in graph theory regarding the zero forcing sets of graphs.
For instance, in case where the underlying graph has a structure of a (directed) tree, we conclude that the minimum number of leaders rendering the network controllable, for all matrices in the qualitative class, is equal to the corresponding path cover number of the graph.
Moreover, initial vertices in a minimal path cover can be selected as the choice of leaders in this case.
Likewise, one can draw similar conclusions for other classes of graphs for which the zero forcing sets has been already studied in the literature.
Finally, thanks to the result of the present paper, the problem of verifying whether a given set of leaders render the network controllable, for all matrices in the qualitative class, boils down to checking whether this leader set constitutes a zero forcing set or not.

The organization of the paper is as follows.
In Section \ref{ZFS:s:Motivation}, the problem at hand is mathematically formulated, and is motivated by establishing connection to the existing results in the literature.
In Section \ref{ZFS:s:zfs}, zero forcing sets, zero forcing number, and the involved notions are recapped. The main result of the paper is reported in Section \ref{ZFS:s:zfs-controllability}, where a necessary and sufficient condition for controllability of networks is established in terms of zero forcing sets. In addition, controllability of the network with respect to qualitative subclasses is studied in this section, and
finally some special cases are provided for further illustration of the proposed results.
The paper ends with concluding remarks in Section \ref{ZFS:s:conclusion}.

\section{Problem formulation and motivation}\label{ZFS:s:Motivation}
For a given simple directed graph $G$, the vertex set of $G$ is a nonempty set and is denoted by $V(G)$. The arc set of $G$, denoted by $E(G)$, is a subset of $V \times V$, and
$(i,i) \notin E$ for all $i \in V(G)$. The cardinality of a given set $V$ is denoted by $|V|$. Also we use $|G|$ to denote in short the cardinality of $V(G)$.
We say vertex $j$ is an out-neighbor of vertex $i$ if $(i,j) \in E$.
The family of matrices described by $G$ is called {\em qualitative class} of $G$, and is given by

\beq\label{ZFS:Q(G)}
\mathcal{Q}(G)=\{X \in \mathbb{R}^{|G| \times |G|}:\text{ for } i\neq j,\;  X_{ij}\neq 0 \Leftrightarrow (j,i) \in E(G)\}.
\eeq

For $V=\nset{n}$ and $V_L=\pset{v_1,v_2,\ldots,v_m}\subseteq V$, we define the $n\times m$ matrix $U(V;V_L)=[U_{ij}]$ by:
\beq\label{ZFS:U}
U_{ij}=\begin{cases} 1&\text{if }i=v_j\\ 0&\text{otherwise}.\end{cases}
\eeq
By a leader/follower system defined on a graph $G$, we mean a finite-dimensional linear input/state system of the form
\beq\label{ZFS:e:sys-graph}
\dot{x}(t)=Xx(t)+Uu(t)
\eeq
in continuous-time and
\beq\label{ZFS:e:sys-graph-disc}
x(t+1)=Xx(t)+Uu(t)
\eeq
in discrete-time where $x\in\R^{|G|}$ is the state, $u\in\R^m$ is the input, $X\in\calQ(G)$, and $U=U(V(G);V_L)$ for some given leader set $V_L\subseteq V(G)$.\\

Systems of the form \eqref{ZFS:e:sys-graph} or \eqref{ZFS:e:sys-graph-disc} where $X\in\calQ(G)$ for a given graph $G$ are encountered in various contexts. Examples include the cases where $X$ is adjacency \cite{CDG10},  (in-degree or out-degree) Laplacian \cite{MMME10}, normalized Laplacian \cite{ABJJ07}, etc. matrices associated to a graph.

In this paper, we deal with the controllability of the systems of the form \eqref{ZFS:e:sys-graph} or \eqref{ZFS:e:sys-graph-disc}. With a slight abuse of notation, we sometimes write $(X;V_L)$ is controllable meaning that $(X,U)$ is controllable. For a given graph $G$ and a leader set $V_L$ we say $(G;V_L)$ is controllable if the pair $(X;V_L)$ is controllable for all $X \in \calQ(G)$.

In particular, we are interested in determining the set of leaders rendering systems of the form \eqref{ZFS:e:sys-graph} controllable. For a given graph $G$ and a matrix $X \in \calQ(G)$, we denote the minimum number of leaders rendering the system \eqref{ZFS:e:sys-graph} controllable by $\lmin(X)$, that is
$$
\lmin(X)=\min_{V_L\subseteq V(G)}\{|V_L|: (X;V_L) \textrm{ is controllable}\}.
$$

For a given graph $G$, we denote the minimum number of leaders rendering all systems of the form \eqref{ZFS:e:sys-graph} controllable by $\lmin(G)$, that is
\beq\label{ZFS:lminG}
\lmin(G)=\min_{V_L\subseteq V(G)}\{|V_L|: (G;V_L) \textrm{ is controllable}\}.
\eeq

Controllability of systems of the form \eqref{ZFS:e:sys-graph} has been studied in the literature from different angles. In what follows, we give an account of the existing results/approaches in the literature.

One particular line of research within the context of controllability has been devoted to systems of the form
\beq\label{ZFS:e:sys-laplacian}
\dot{x}(t)=-Lx(t)+Uu(t)
\eeq
where $L$ is the Laplacian matrix of an undirected graph. This line of research has been initiated by \cite{HGT04} and further developed by \cite{ARMJMMME09}. Within this framework, the two main themes were graph theoretical characterization of controllability properties in terms of certain graph partitions \cite{SMMEAB10}, \cite{MESMMCMKCAB12}, \cite{SZMKCMC11} and (minimum) leader selection for rendering a system of the form \eqref{ZFS:e:sys-laplacian} controllable for particular classes of undirected graphs \cite{SZMKCMC11}, \cite{GPGN12}, \cite{MNMMta}.

The work on the leader selection led to a number of interesting results by exploiting the structure of the Laplacian matrices for several graph classes. It has been shown in \cite{ARMJMMME09} that $\lmin(L)=1$ for path graphs. In this case, one can choose one of the two terminal vertices as the leader. By \cite{SZMKCMC11}, $\lmin(L)=2$ for undirected cycle graphs and any two neighbours can be chosen as leaders. The paper \cite{GPGN12} further studied cycle graphs and has proved that any two leaders would render the system controllable in case the number of all vertices is a prime number. For an undirected complete graph with $n$ vertices, we know from \cite{SZMKCMC11}, \cite{ARMJMMME09} that $\lmin(L)=n-1$ and any choice of $n-1$ leaders would render the system controllable. Another rather specific class of undirected graphs that has been studied within the same context is distance regular graphs. In \cite{SZMKCMC11}, it was shown that $\lmin(L)\leq n-d$ where $n$ is the number of vertices and $d$ is the diameter of the graph. The paper \cite{SZMKCMC11} provided also a recipe to select $n-d$ leaders that render the system controllable. In case the underlying graph is a circulant graph, the authors of \cite{MNMMta} proved that $\lmin(L)$ is equal to the maximum algebraic multiplicity of Laplacian eigenvalues.

Another particular class of systems that has been studied in the context of the controllability is given by
\beq\label{ZFS:e:sys-adjacency}
\dot{x}(t)=Ax(t)+Uu(t)
\eeq
where $A$ is the adjacency matrix of an undirected graph, see e.g. \cite{CDG10}. The same class of systems was studied in \cite{YYLJJSALB11} from the weak structural controllability viewpoint.

In this paper, we will mainly deal with the controllability of families of systems given by \eqref{ZFS:e:sys-graph} where $X\in\calQ(G)$ for a graph $G$ and provide results concerning $\lmin(G)$ rather than $\lmin(X)$ for a specific choice of $X\in\calQ(G)$. However, our treatment, as a side result, will reveal that the aforementioned existing results on the number of minimum leaders are not intrinsic to the Laplacian but hold for any matrix within the corresponding qualitative class given by the underlying graph.

\section{Zero forcing sets}\label{ZFS:s:zfs}

First, we review the notion of zero forcing sets together with the involved notations and terminology which will be used in the sequel.
For more details see e.g. \cite{LH10}.

Let $G$ be a given graph, where each vertex is colored either white or black. Consider the following coloring rule:\\[-2mm]

$\LEFTcircle:$ If $u$ is a black vertex and exactly one out-neighbor $v$ of $u$ is
white, then change the color of $v$ to black.

Following terminology will be used when we apply the color-change rule above to a graph $G$:

\bit
\item[--] When the color-change rule is applied to $u\in V(G)$ to change the color of $v \in V(G)$, we say $u$ {\em forces} or {\em infects} $v$, and write $u \rightarrow v$.

\item[--] Given a coloring set $C \subseteq V(G)$, i.e. $C$ indexes the initially black vertices of $G$, the {\em derived\/} set of $C$ is denoted by $\calD(C)$, and is the set of black vertices obtained by applying the color-change rule until no more changes are possible.

\item[--] The set $Z \subseteq V(G)$ is a {\em zero forcing set\/} (ZFS) for $G$ if $\calD(Z)=V(G)$.

\item[--] The {\em zero forcing number} $Z(G)$ is the minimum of $|Z|$ over all zero forcing sets $Z \subseteq V (G)$. A set $Z$ is called a {\em minimal zero forcing set\/}
if $|Z|=Z(G)$.
\eit

Figures~\ref{ZFS:f:NOTzfs} and~\ref{ZFS:f:zfs} illustrate the zero forcing set and the notions defined above. First, consider the graph depicted
in Figure \ref{ZFS:f:NOTzfs} where the vertex $1$ is initially colored black.
Then, by the color-change rule it is clear that $1 \rightarrow 2$. Consequently, $2 \rightarrow 3$, and $3 \rightarrow 4$. Therefore, the derived set of $\{1\}$ is equal to $\{1, 2, 3, 4\}$, and thus
$\{1\}$ is not a zero forcing set. Now, suppose that we choose $\{1, 5\}$ to be the initially colored black vertices as shown in Figure~\ref{ZFS:f:zfs}. Then by applying the color-change rule,
we conclude that this set is a zero forcing set. Moreover, note that no singleton set constitutes a zero forcing set in this case, thus the zero forcing number is indeed equal to 2.

\begin{figure}[ht!]
\centering
\begin{minipage}{0.5 \textwidth}
\[\begin{tikzpicture}[x=.6cm, y=.6cm,
    every edge/.style={
        draw,
        postaction={decorate,
                    decoration={markings,mark=at position 0.6 with {\arrow{>}}}
                   }
        }
]

\vertex (v1)[ball color=black!100!] at (-5,0) [label=above:$1$] {};
\vertex (v2) at (-4,0) [label=above:$2$] {};
\vertex (v3) at (-3,0) [label=above:$3$] {};
\vertex (v4) at (-3,-1) [label=below:$4$] {};
\vertex (v5) at (-4,-1) [label=below:$5$] {};
\vertex (v6) at (-5,-1) [label=below:$6$] {};

   \path
		(v1) edge (v2)
		(v2) edge (v3)
        (v3) edge[bend right,in=135,out=45] (v4)
        (v4) edge[bend right,in=135,out=45] (v3)
        (v5) edge (v4)
        (v5) edge (v6)
        (v5) edge (v1)
        (v5) edge (v2)
        (v6) edge (v1)
        ;
\vertex (v1)[ball color=black!100!] at (0,0) [label=above:$1$] {};
\vertex (v2)[ball color=black!100!] at (1,0) [label=above:$2$] {};
\vertex (v3) at (2,0) [label=above:$3$] {};
\vertex (v4) at (2,-1) [label=below:$4$] {};
\vertex (v5) at (1,-1) [label=below:$5$] {};
\vertex (v6) at (0,-1) [label=below:$6$] {};

    \path
		(v1) edge (v2)
		(v2) edge (v3)
        (v3) edge[bend right,in=135,out=45] (v4)
        (v4) edge[bend right,in=135,out=45] (v3)
        (v5) edge (v4)
        (v5) edge (v6)
        (v5) edge (v1)
        (v5) edge (v2)
        (v6) edge (v1)
        ;
\vertex (v1)[ball color=black!100!] at (5,0) [label=above:$1$] {};
\vertex (v2)[ball color=black!100!] at (6,0) [label=above:$2$] {};
\vertex (v3)[ball color=black!100!] at (7,0) [label=above:$3$] {};
\vertex (v4) at (7,-1) [label=below:$4$] {};
\vertex (v5) at (6,-1) [label=below:$5$] {};
\vertex (v6) at (5,-1) [label=below:$6$] {};

    \path
		(v1) edge (v2)
		(v2) edge (v3)
        (v3) edge[bend right,in=135,out=45] (v4)
        (v4) edge[bend right,in=135,out=45] (v3)
        (v5) edge (v4)
        (v5) edge (v6)
        (v5) edge (v1)
        (v5) edge (v2)
        (v6) edge (v1)
        ;

\vertex (v1)[ball color=black!100!] at (-5,-3) [label=above:$1$] {};
\vertex (v2)[ball color=black!100!] at (-4,-3) [label=above:$2$] {};
\vertex (v3)[ball color=black!100!] at (-3,-3) [label=above:$3$] {};
\vertex (v4)[ball color=black!100!] at (-3,-4) [label=below:$4$] {};
\vertex (v5) at (-4,-4) [label=below:$5$] {};
\vertex (v6) at (-5,-4) [label=below:$6$] {};

   \path
		(v1) edge (v2)
		(v2) edge (v3)
        (v3) edge[bend right,in=135,out=45] (v4)
        (v4) edge[bend right,in=135,out=45] (v3)
        (v5) edge (v4)
        (v5) edge (v6)
        (v5) edge (v1)
        (v5) edge (v2)
        (v6) edge (v1)
        ;

\end{tikzpicture}\]
\caption{An example for the coloring rule}
\label{ZFS:f:NOTzfs}
\end{minipage}

\begin{minipage}{0.5 \textwidth}
\centering

\[\begin{tikzpicture}[x=0.6cm, y=0.6cm,
    every edge/.style={
        draw,
        postaction={decorate,
                    decoration={markings,mark=at position 0.6 with {\arrow{>}}}
                   }
        }
]

\vertex (v1)[ball color=black!100!] at (-5,0) [label=above:$1$] {};
\vertex (v2) at (-4,0) [label=above:$2$] {};
\vertex (v3) at (-3,0) [label=above:$3$] {};
\vertex (v4) at (-3,-1) [label=below:$4$] {};
\vertex (v5)[ball color=black!100!] at (-4,-1) [label=below:$5$] {};
\vertex (v6) at (-5,-1) [label=below:$6$] {};

   \path
		(v1) edge (v2)
		(v2) edge (v3)
        (v3) edge[bend right,in=135,out=45] (v4)
        (v4) edge[bend right,in=135,out=45] (v3)
        (v5) edge (v4)
        (v5) edge (v6)
        (v5) edge (v1)
        (v5) edge (v2)
        (v6) edge (v1)
        ;
\vertex (v1)[ball color=black!100!] at (0,0) [label=above:$1$] {};
\vertex (v2)[ball color=black!100!] at (1,0) [label=above:$2$] {};
\vertex (v3) at (2,0) [label=above:$3$] {};
\vertex (v4) at (2,-1) [label=below:$4$] {};
\vertex (v5)[ball color=black!100!] at (1,-1) [label=below:$5$] {};
\vertex (v6) at (0,-1) [label=below:$6$] {};

    \path
		(v1) edge (v2)
		(v2) edge (v3)
        (v3) edge[bend right,in=135,out=45] (v4)
        (v4) edge[bend right,in=135,out=45] (v3)
        (v5) edge (v4)
        (v5) edge (v6)
        (v5) edge (v1)
        (v5) edge (v2)
        (v6) edge (v1)
        ;
\vertex (v1)[ball color=black!100!] at (5,0) [label=above:$1$] {};
\vertex (v2)[ball color=black!100!] at (6,0) [label=above:$2$] {};
\vertex (v3)[ball color=black!100!] at (7,0) [label=above:$3$] {};
\vertex (v4) at (7,-1) [label=below:$4$] {};
\vertex (v5)[ball color=black!100!] at (6,-1) [label=below:$5$] {};
\vertex (v6) at (5,-1) [label=below:$6$] {};

    \path
		(v1) edge (v2)
		(v2) edge (v3)
        (v3) edge[bend right,in=135,out=45] (v4)
        (v4) edge[bend right,in=135,out=45] (v3)
        (v5) edge (v4)
        (v5) edge (v6)
        (v5) edge (v1)
        (v5) edge (v2)
        (v6) edge (v1)
        ;

\vertex (v1)[ball color=black!100!] at (-5,-3) [label=above:$1$] {};
\vertex (v2)[ball color=black!100!] at (-4,-3) [label=above:$2$] {};
\vertex (v3)[ball color=black!100!] at (-3,-3) [label=above:$3$] {};
\vertex (v4)[ball color=black!100!] at (-3,-4) [label=below:$4$] {};
\vertex (v5)[ball color=black!100!] at (-4,-4) [label=below:$5$] {};
\vertex (v6) at (-5,-4) [label=below:$6$] {};

   \path
		(v1) edge (v2)
		(v2) edge (v3)
        (v3) edge[bend right,in=135,out=45] (v4)
        (v4) edge[bend right,in=135,out=45] (v3)
        (v5) edge (v4)
        (v5) edge (v6)
        (v5) edge (v1)
        (v5) edge (v2)
        (v6) edge (v1)
        ;

\vertex (v1)[ball color=black!100!] at (0,-3) [label=above:$1$] {};
\vertex (v2)[ball color=black!100!] at (1,-3) [label=above:$2$] {};
\vertex (v3)[ball color=black!100!] at (2,-3) [label=above:$3$] {};
\vertex (v4)[ball color=black!100!] at (2,-4) [label=below:$4$] {};
\vertex (v5)[ball color=black!100!] at (1,-4) [label=below:$5$] {};
\vertex (v6)[ball color=black!100!] at (0,-4) [label=below:$6$] {};

   \path
		(v1) edge (v2)
		(v2) edge (v3)
        (v3) edge[bend right,in=135,out=45] (v4)
        (v4) edge[bend right,in=135,out=45] (v3)
        (v5) edge (v4)
        (v5) edge (v6)
        (v5) edge (v1)
        (v5) edge (v2)
        (v6) edge (v1)
        ;

\end{tikzpicture}\]
\caption{An example for the zero forcing set}
\label{ZFS:f:zfs}
\end{minipage}
\end{figure}

\section{Zero forcing sets and controllability}\label{ZFS:s:zfs-controllability}

In this section, we characterize a set of leaders which renders $(G;V_L)$ controllable for a given graph $G$. Clearly, a pair $(X,U)$ is controllable if and only if the matrix $\bbm X-\lambda I& U\ebm$ has full row rank for all $\lambda \in \mathbb{C}$.
Here, we deal with a family of matrices based on a given graph $G$, and thus we should consider whether the matrix $\bbm X-\lambda I& U\ebm$ has full row rank for all $X \in \calQ(G)$
and $\lambda \in \mathbb{C}$. It turns out that this property does not depend on the parameter $\lambda$ due to the structure of the matrix family $\calQ(G)$.

\begin{lemma}\label{ZFS:l:pbh}
Let $G$ be a graph and $V_L\subseteq V(G)$. Then, $(G;V_L)$ is controllable if and only if the matrix $\bbm X & U \ebm$ has full row rank for all $X \in \mathcal{Q}(G)$ where $U=U(V;V_L)$ given
by \eqref{ZFS:U}.
\end{lemma}

\BP
Clearly, $(G;V_L)$ is controllable if and only if the matrix $\bbm X-\lambda I & U \ebm$ has full row rank for all $X \in \mathcal{Q}(G)$ and all $\lambda \in \mathbb{C}$.
Hence, the ``only if'' part follows trivially. Now, suppose that $\bbm X & U\ebm$ has full row rank for all $X \in \calQ(G)$. Let $\lambda \in \mathbb{C}$ and $z \in \mathbb{C}^{|G|}$ be such that $z^\ast \bbm X-\lambda I & U\ebm=0$ for some $X \in \calQ(G)$. Let $z=p+ jq$ for real vectors $p$ and $q$ where $j$ is the imaginary number. Define $x \in \mathbb{R}^{|G|}$ as $x=p+\alpha q$ where $\alpha$ is a real number.
Choose $\alpha$ such that
\beq\label{ZFS:alpha}
\alpha \notin \{ -\frac{p_i}{q_i}: q_i \neq 0; i=1, 2, \ldots, |G| \}
\eeq
where $p_i$ and $q_i$ denote the $i^{th}$ element of $p$ and $q$, respectively.
Then one can show that $x_i=0$ if and only if $z_i=0$.
In fact, if $z_i=0$ then obviously $x_i=0$. In addition, if $x_i=0$ then we obtain $p_i+\alpha q_i=0$, which yields $q_i=0$ by \eqref{ZFS:alpha}.
Hence, we have $p_i=0$, and thus $z_i=0$.

Next, we claim that the following implication holds:
\beq\label{ZFS:implication}
x_i=0 \Rightarrow (x^\top X)_i=0.
\eeq
To prove this claim, suppose that $x_i=0$. Then, we have $z_i=0$. Since $z^\ast X=\lambda z^\ast$, we obtain $(z^\ast X)_i=0$. Hence, $(p^\top X)_i=0=(q^\top X)_i$. Consequently, $((p^\top +\alpha q^\top)X)_i=(x^\top X)_i=0$.

Now, we define the diagonal matrix $D=\mathrm{diag}(d_1, d_2, \ldots, d_n)$ with
\beq
d_i=
\begin{cases}
0 &\text{if }x_i=0\\
\frac{(x^\top X)_i}{x_i}&\text{otherwise}.
\end{cases}
\eeq

By using \eqref{ZFS:implication}, it holds that $x^\top X=x^\top D$. Besides, $z^\ast U=0$ results in $p^\top U=0=q^\top U$ which yields $x^\top U=0$. Now, choose $\hat{X}=X-D$. Clearly, $\hat{X} \in \calQ(G)$ and $x^\top \hatX=0$.
Then due to full row rank assumption of $\bbm \hat{X} & U\ebm$
we obtain $x=0$, thus $z=0$. Therefore, $\bbm X-\lambda I & U\ebm$ has full row rank, and the result follows.
\EP

Next, we explore the relationship between zero forcing sets and controllability of $(G;V_L)$.
First we show that the process of coloring/infecting vertices, according to the change-color rule, does not affect the controllability.
This issue is addressed in the following lemma.

\begin{lemma}\label{ZFS:l:color-contr}
Let $G$ be a graph and $C$ be a (coloring) set. Suppose that $v \rightarrow w$ where $v\in C$ and $w \notin C$.
Then $(G;C)$ is controllable if and only if $(G;C \cup \{w\})$
is controllable.
\end{lemma}

\BP
The ``only if" part is trivial. Now, let $C^\prime:=C \cup \{w\}$ and suppose that $(G;C^\prime)$ is controllable. Hence, $(X,U)$ is controllable for all $X \in \calQ(G)$ where $U=U(V(G);C^\prime)$ is given by \eqref{ZFS:U}. Without loss of generality, we can assume that
\beq\label{ZFS:part-1}
(X,U)=\Big(
\bbm
x_{11} &x_{12} &x_{13} &x_{14}\\
x_{21} &x_{22} &x_{23} &x_{24}\\
X_{31} & X_{32}&X_{33} &X_{34}\\
X_{41} & X_{42}&X_{43} &X_{44}
\ebm,
\bbm
1 & 0 &0\\
0 &1 &0\\
0 & 0& I\\
0 &0 &0
\ebm\Big)
\eeq
is controllable for all $X \in \calQ(G)$, where the first row corresponds to the vertex $w$, the second corresponds to $v$, the third row block corresponds to the vertices indexed by $C\setminus \{v\}$, and the last row block corresponds to remaining white vertices, i.e. $V(G)\setminus C^\prime$.
By Lemma \ref{ZFS:l:pbh}, we know that $\bbm X &U\ebm$ has full row rank, which implies that the last row block of $X$ in \eqref{ZFS:part-1} has full row rank. Since $v \rightarrow w$, we have $x_{12}\neq 0$ and $X_{42}=0$. Therefore, the submatrix
\beq
\bbm
x_{11} &x_{12} &x_{13} &x_{14}\\
X_{41} & X_{42}&X_{43} &X_{44}
\ebm
\eeq
has full row rank.
Consequently, the pair
\beq\label{ZFS:part-2}
\Big(
\bbm
x_{11} &x_{12} &x_{13} &x_{14}\\
x_{21} &x_{22} &x_{23} &x_{24}\\
X_{31} & X_{32}&X_{33} &X_{34}\\
X_{41} & X_{42}&X_{43} &X_{44}
\ebm,
\bbm
0 & 0 &0\\
0 &1 &0\\
0 & 0& I\\
0 &0 &0
\ebm\Big)
\eeq
is controllable, and hence $(G;C)$ is controllable.
\EP

Roughly speaking, this lemma states that controllability is invariant under infection. As such, we can obtain the following corollary by repeated application of Lemma \ref{ZFS:l:color-contr}.

\begin{corollary}\label{ZFS:c:derivedset}
Let $G$ be a graph and a $C$ be a coloring set. Then, $(G;C)$ is controllable if and only if $(G;\calD(C))$ is controllable.
\end{corollary}

Next, we state one of the main results of the paper based on the above auxiliary lemmas.

\begin{theorem}\label{ZFS:t:zfs-contr-directed}
Let $G$ be a graph and $V_L \subseteq V(G)$. Then, $(G;V_L)$ is controllable if and only if $V_L$ is a zero forcing set.
\end{theorem}

\BP
If $V_L$ is a zero forcing set, then $\calD(V_L)=V(G)$ by definition. Hence, it follows from
Corollary~\ref{ZFS:c:derivedset} that controllability of $(G;V_L)$ is equivalent to that
of $(G;V(G))$. Since $(G;V(G))$ is trivially controllable, so is $(G;V_L)$. To prove the converse, suppose that $(G;V_L)$ is controllable, but $V_L$ is {\em not} a zero forcing set.
Then, we have $\calD(V_L) \neq V(G)$. We also know that $(G;\calD(V_L))$ is controllable by Corollary \ref{ZFS:c:derivedset}. Without loss of generality, we can assume that $V_L=\{1,2,...,m\}$ and $\calD(V_L)=V_L \cup \{m+1, m+2, \ldots, m+r\}$ where $m+r < |G|$.
Since $(G;\calD(V_L))$ is controllable, it follows from Lemma \ref{ZFS:l:pbh} that the matrix $\bbm X & U\ebm$ has full row rank for all $X \in \calQ(G)$ where $U=U(V;\calD (V_L))=\mathrm{col}(I_{m+r}, 0)$.
Hence, the matrix
\beq\label{ZFS:part-3}
\bbm
X_{11} & X_{12} & I_{m+r}\\
X_{21} & X_{22} & 0
\ebm
\eeq
has full row rank for all $X \in \calQ(G)$ where $X_{11}\in\R^{(m+r)\times(m+r)}$, $X_{12}\in\R^{(m+r)\times k}$, $X_{21}\in\R^{k \times(m+r)}$, and $X_{22}\in\R^{k\times k}$ with $k=\abs{G}-(m+r)$ constitute the corresponding partitioning of the matrix $X$.

Now, we distinguish two cases. First, suppose that there exists a column of $X_{21}$ with exactly one nonzero element.
This implies that there is a vertex, say $v \in \calD(V_L)$, which has exactly one (white) out-neighbor, say $w \notin \calD(V_L)$. Consequently, $v$ can infect $w$, and we reach a contradiction.
On the other hand, suppose that there does not exist a column of $X_{21}$ with exactly one nonzero element.
Then, clearly the nonzero elements of $X_{21}$ can be chosen such that we have $\ones ^\top X_{21}=0$, where $\ones$ denotes the vector of ones with an appropriate dimension.
In addition, note that the diagonal elements of $X$ can be chosen arbitrarily due to the the definition of $\calQ(G)$, and thus can be assigned such that $\ones ^\top X_{22}=0$.
Therefore, we obtain that
$$\bbm 0_{m+r}^\top & \ones^\top \ebm  \bbm
X_{11} & X_{12} & I_{m+r}\\
X_{21} & X_{22} & 0
\ebm =0,$$
for some $X$ in $Q(G)$, and again we reach a contradiction.
\EP
\bre\label{ZFS:r:input-connectivity}
In case $V_L$ is a zero forcing set, it is easy to observe that each vertex of $V\setminus V_L$ is accessible (via a directed path) from at least one leader.
 This input-accessibility condition is indeed necessary for weak/strong structural controllability of networks
(see e.g. \cite{CTL74} and \cite[Thm. 1]{YYLJJSALB11}).
\ere

Theorem \ref{ZFS:t:zfs-contr-directed} establishes a one-to-one correspondence between leader sets rendering systems of the form \eqref{ZFS:e:sys-graph} controllable and zero forcing sets of the corresponding graphs. An immediate consequence of this result yields the following result on the minimum number of leaders required for controllability.
\begin{corollary}\label{ZFS:c:lmin-zfn}
Let $G$ be a given graph. Then, $\lmin(G)=Z(G)$.
\end{corollary}

\subsection{Sufficiently rich qualitative subclasses}\label{ZFS:ss:rich}

So far, we have investigated controllability of systems given by \eqref{ZFS:e:sys-graph} where the matrices $X$ belongs to the family $\calQ(G)$ which is described by the graph $G$. In many examples, one encounters matrices of $X$ carrying more structure than that is imposed by $\calQ(G)$. For instance, consider a graph $G_1$ for which $E(G_1)$ is symmetric, i.e. $(v,w) \in E(G_1)$ if and only if $(w,v) \in E(G_1)$ and the matrices $X$ belonging to
\beq\label{ZFS:e:g1}
\Qyek=\{X\in\calQ(G_1):X=X^\top\}\subseteq\calQ(G_1).
\eeq
Note that undirected graphs can be identified with directed graphs having symmetric arc sets. As such, the class $\Qyek$ naturally appears whenever the underlying graph structure is induced by an undirected graph as in the systems of the form \eqref{ZFS:e:sys-laplacian} and \eqref{ZFS:e:sys-adjacency}

In what follows, we focus on controllability with respect to subclasses of $\calQ(G)$. For a graph $G$, (leader) set $V_L\subseteq V(G)$, and a
qualitative subclass $\calQ^\prime(G) \subseteq \calQ(G)$, we say $V_L$ controls $\calQ^\prime(G)$ if $(X;V_L)$ is controllable for all $X \in \calQ^\prime (G)$.

If $V_L$ is a zero a forcing set for the graph $G$, then $V_L$ controls $\calQ(G)$ by Theorem~\ref{ZFS:t:zfs-contr-directed}. Consequently, such a $V_L$ controls $\calQ^\prime (G)$ for any $\calQ^\prime(G) \subseteq \calQ(G)$. However, the converse is not true in general.
For instance, consider $G_1=(V_1,E_1)$ where $V_1=\{1,2,3,4\}$ and $E_1=\{(1,2),(2,1),(2,3),(3,2),(3,4),(4,3)\}$. Let $V_L=\{2\}$ and take the Laplacian matrix of $G_1$, denoted by $L_1$, as the qualitative subclass in this case. Then, by \cite{GPGN12}, $(L_1;V_L)$ is controllable whereas obviously $V_L$ is not a zero forcing set.

%

Therefore, we conclude that $V_L$ is not necessarily a zero forcing set for $G$ even though it controls a nonempty subset of
$\calQ(G)$. Next, we investigate under what conditions, controlling a subset of $\calQ(G)$ implies that $V_L$ is a zero forcing set. For this purpose, the following notion is needed.

\begin{definition}
Let $\calQ^\prime(G)$ be a non-empty subset of $\calQ(G)$. We say that $\calQ^\prime(G)$ is a {\em sufficiently rich} subclass of $\calQ(G)$ if the following implication holds:
\beq\label{ZFS:suff-rich}
z \in \mathbb{R}^{|G|},\,X\in \calQ(G),\,z^T X=0
\;\Longrightarrow\; \exists\, X^\prime \in \calQ^\prime(G) \textrm{ s.t. } z^T X^\prime=0.
\eeq
\end{definition}

Now, we have the following result.
\begin{theorem}\label{ZFS:t:rich}
Let $G$ be a graph and $V_L\subseteq V(G)$ be a (leader) set. Suppose that $\calQ^\prime (G) \subseteq \calQ(G)$ is a sufficiently rich subclass of $\calQ(G)$. Then the following statements are equivalent:
\ben
\item The set $V_L$ is a zero forcing set.
\item The set $V_L$ controls $\calQ(G)$.
\item The set $V_L$ controls $\calQ^\prime(G)$.
\een
\end{theorem}

\BP
The first two statements are equivalent by Theorem \ref{ZFS:t:zfs-contr-directed}. Besides, the second statement trivially implies the third one.
Hence, it suffices to show that statement $3$ implies $2$. Suppose that statement $3$ holds. In view of Lemma \ref{ZFS:l:pbh}, it suffices to show that
the matrix $\bbm X &U\ebm$ has full row rank for all $X \in \calQ(G)$, where $U$ is given by \eqref{ZFS:U}. Now suppose that $x^\top \bbm X &U\ebm=0$ for some
$x\in \mathbb{R}^{|G|}$ and $X \in \calQ(G)$. Since $\calQ^\prime(G)$ is a sufficiently rich subclass of $\calQ(G)$, there exists $X^\prime \in \calQ^\prime(G)$ such that
$x^\top \bbm X^\prime &U\ebm=0$. This results in $x=0$ due to the assumption that $V_L$ controls $\calQ^\prime(G)$.  Consequently, the matrix $\bbm X &U\ebm$ has full row rank for all $X \in \calQ(G)$. Thus, $V_L$ also controls $\calQ(G)$.
\EP

By Theorem \ref{ZFS:t:rich}, controlling sufficiently rich subclasses is equivalent to controlling the corresponding qualitative classes, which can be further characterized by zero forcing sets.
Next, we focus on two notable subclasses of $\calQ(G_1)$. Bare in mind that $E(G_1)$ is symmetric. The first subclass we consider here is $\Qyek$ given by \eqref{ZFS:e:g1}.

\begin{proposition}\label{ZFS:p:Qyek}
The set $\Qyek$ is a sufficiently rich subclass of $\calQ(G_1)$.
\end{proposition}

\BP
Assume that there exists $z\in \mathbb{R}^{|G_1|}$ such that $z^\top X=0$ for some $X \in \calQ(G_1)$.
We distinguish two cases. First, suppose that $z_i \neq 0$ for each $i=1, 2, \ldots, |G_1|$. Define the matrix $X'$ as $X'=\hatX+D$ where $\hatX\in \Qyek$ and $D$ is a real diagonal matrix.
Obviously, we have $X' \in \Qyek$ for any choice of $D$. Then, since $z_i\neq 0$ for each $i$, one can choose $D$ such that $z^\top X'=0$.
Next, consider the case where $z_i=0$ for some $i$. Without loss of generality, the vector $z$ can be then decomposed as $z=[\hat{z}^\top \quad 0] ^\top $
such that the vector $\hat{z}$ does not contain any zero element.
Correspondingly, let the matrix $X$ be decomposed as
$$
X=
\bbm
X_{11} & X_{12}\\
X_{21}  & X_{22}
\ebm.
$$
Hence, we have $\hat{z}^\top X_{11}=0$ and $\hat{z}^\top X_{12}=0$ by the assumption. Now, choose a matrix $\hat{X}\in \Qyek$ and let
$$
\hatX=
\bbm
\hatX_{11}  & \hatX_{12}\\
\hatX_{12}^\top & \hatX_{22}
\ebm.
$$
Let $D$ be a real diagonal matrix such that $\hat{z}^\top (\hatX_{11}+D)=0$. Note that such $D$ exists as $\hat{z}_i\neq 0$ for each $i$.
Then, we construct a matrix $X'$ as
$$
X'=
\bbm
\hatX_{11}+D  & X_{12}\\
X_{12}^\top & \hatX_{22}
\ebm.
$$
Clearly, we have $X' \in \Qyek$. Moreover, it holds that $z^\top X'=0$, and thus $\Qyek$ is a sufficiently rich subclass of $\calQ(G_1)$.
\EP

Now, we consider another subclass of $\calQ(G_1)$
by imposing an additional constraint to $\Qyek$.
More precisely, let $\Qdo$ be defined as a subset of $\Qyek$ with the property that all off-diagonal nonzero elements of $X$ have the same sign for all $X \in \Qdo$.
Note that ordinary Laplacian matrices and adjacency matrices are among the special cases of this subclass.
Structural controllability with respect to $\Qdo$ has been studied in \cite{DBDDLHSSMYta}. In particular, it has been shown that the set $V_L$ controls $\Qdo$ if $V_L$ is a zero forcing set.
However, the converse does not hold in general (see \cite[Ex. 4.3]{DBDDLHSSMYta}). The following proposition shows that indeed $\Qdo$ is not a sufficiently rich subclass, except for some pathological cases.

\begin{proposition}\label{ZFS:p:Qdo}
Assume that the graph $G_1$ has a vertex with at least two (out) neighbors. Then, the set $\Qdo$ is not a sufficiently rich subclass of $\calQ(G_1)$.
\end{proposition}

 \BP
 Let $k$ be a vertex of $G_1$ with at least two (out) neighbors. Define $z\in \R^{|G_1|}$ as
 $$
 z_i=
 \begin{cases}
  1  &\textrm{ if } i \neq k,\\
  0  & \textrm{ otherwise. }
 \end{cases}
 $$
Note that $(z^\top X')_k$ is nonzero for any $X'\in \Qdo.$ Hence, $z^\top X'\neq 0$ for any $X'\in \Qdo.$ Therefore, to conclude that $\Qdo$ is not sufficiently rich, it suffices to show that $z^\top X=0$ for some $X\in\calQ(G_1)$.
It is easy to see that one can choose a matrix $X \in \calQ(G_1)$ such that $(z^\top X)_i=0$ for each $i\neq k$.
Also note that, by the assumption, the matrix $X$ has at least two nonzero off-diagonal elements in its $k^{th}$ column.
Hence, these (two or more) nonzero elements can be further chosen such that we have $(z^\top X)_k=0$, and thus $z^\top X=0$. This completes the proof.
 \EP

\subsection{Special cases}
Next, we study some special cases to demonstrate how the proposed results can be used in particular applications.

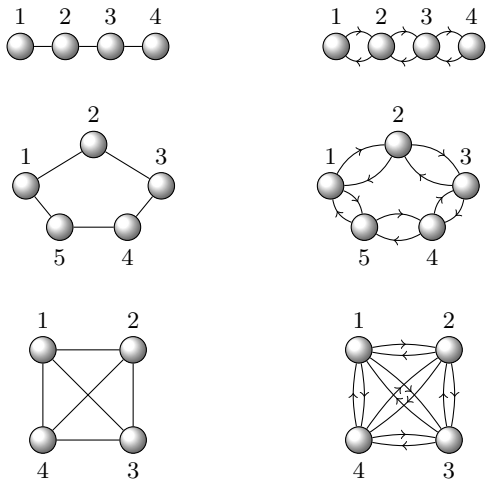
\begin{figure}

\begin{minipage}{0.5 \textwidth}


\[\begin{tikzpicture}[x=.6cm, y=.6cm,
    every edge/.style={draw}]



\vertex (v1) at (-1,0) [label=above:$1$] {};
\vertex (v2) at (0,0) [label=above:$2$] {};
\vertex (v3) at (1,0) [label=above:$3$] {};
\vertex (v4) at (2,0) [label=above:$4$] {};

\path
		(v1) edge (v2)
		(v2) edge (v3)
        (v3) edge (v4)
                ;

\vertex (v1) at (6,0) [label=above:$1$] {};
\vertex (v2) at (7,0) [label=above:$2$] {};
\vertex (v3) at (8,0) [label=above:$3$] {};
\vertex (v4) at (9,0) [label=above:$4$] {};

\path
		(v1) edge[postaction={decorate,
                   decoration={markings,mark=at position 0.6 with {\arrow{>}}}},bend right,in=135,out=45] (v2)
		(v2) edge[postaction={decorate,
                   decoration={markings,mark=at position 0.6 with {\arrow{>}}}},bend right,in=135,out=45] (v3)
        (v3) edge[postaction={decorate,
                   decoration={markings,mark=at position 0.6 with {\arrow{>}}}},bend right,in=135,out=45] (v4)
        (v4) edge[postaction={decorate,
                   decoration={markings,mark=at position 0.6 with {\arrow{>}}}},bend right,in=135,out=45] (v3)
		(v3) edge[postaction={decorate,
                   decoration={markings,mark=at position 0.6 with {\arrow{>}}}},bend right,in=135,out=45] (v2)
        (v2) edge[postaction={decorate,
                   decoration={markings,mark=at position 0.6 with {\arrow{>}}}},bend right,in=135,out=45] (v1)

                ;
\end{tikzpicture}\]
\end{minipage}
\bigskip

\begin{minipage}{0.5\textwidth}
\[\begin{tikzpicture}[x=.45cm, y=.55cm,
    every edge/.style={draw}]

\vertex (v1) at (0,1) [label=above:$1$] {};
\vertex (v2) at (2,2) [label=above:$2$] {};
\vertex (v3) at (4,1) [label=above:$3$] {};
\vertex (v4) at (3,0) [label=below:$4$] {};
\vertex (v5) at (1,0) [label=below:$5$] {};

\path
		(v1) edge (v2)
		(v2) edge (v3)
        (v3) edge (v4)
        (v4) edge (v5)
        (v5) edge (v1)

        ;

\vertex (v1) at (9,1) [label=above:$1$] {};
\vertex (v2) at (11,2) [label=above:$2$] {};
\vertex (v3) at (13,1) [label=above:$3$] {};
\vertex (v4) at (12,0) [label=below:$4$] {};
\vertex (v5) at (10,0) [label=below:$5$] {};

\path
		(v1) edge[postaction={decorate,
                   decoration={markings,mark=at position 0.6 with {\arrow{>}}}},bend right,in=150,out=30] (v2)
		(v2) edge[postaction={decorate,
                   decoration={markings,mark=at position 0.6 with {\arrow{>}}}},bend right,in=150,out=30] (v3)
        (v3) edge[postaction={decorate,
                   decoration={markings,mark=at position 0.6 with {\arrow{>}}}},bend right,in=150,out=30] (v4)
        (v4) edge[postaction={decorate,
                   decoration={markings,mark=at position 0.6 with {\arrow{>}}}},bend right,in=150,out=30] (v5)
        (v5) edge[postaction={decorate,
                   decoration={markings,mark=at position 0.6 with {\arrow{>}}}},bend right,in=150,out=30] (v1)
        (v1) edge[postaction={decorate,
                   decoration={markings,mark=at position 0.6 with {\arrow{>}}}},bend right,in=150,out=30] (v5)
        (v5) edge[postaction={decorate,
                   decoration={markings,mark=at position 0.6 with {\arrow{>}}}},bend right,in=150,out=30] (v4)
        (v4) edge[postaction={decorate,
                   decoration={markings,mark=at position 0.6 with {\arrow{>}}}},bend right,in=150,out=30] (v3)
        (v3) edge[postaction={decorate,
                   decoration={markings,mark=at position 0.6 with {\arrow{>}}}},bend right,in=150,out=30] (v2)
        (v2) edge[postaction={decorate,
                   decoration={markings,mark=at position 0.6 with {\arrow{>}}}},bend right,in=150,out=30] (v1)
        ;

\end{tikzpicture}\]
\end{minipage}

\bigskip
\begin{minipage}{0.5 \textwidth}

\[\begin{tikzpicture}[x=.6cm, y=.6cm,
    every edge/.style={draw}]

\vertex (v1) at (-1,2) [label=above:$1$] {};
\vertex (v2) at (1,2) [label=above:$2$] {};
\vertex (v3) at (1,0) [label=below:$3$] {};
\vertex (v4) at (-1,0) [label=below:$4$] {};

    \path
		(v1) edge (v2)
		(v2) edge (v3)
        (v3) edge (v4)
		(v4) edge (v1)
		(v1) edge (v3)
        (v2) edge (v4)
                ;

\vertex (v1) at (6,2) [label=above:$1$] {};
\vertex (v2) at (8,2) [label=above:$2$] {};
\vertex (v3) at (8,0) [label=below:$3$] {};
\vertex (v4) at (6,0) [label=below:$4$] {};

    \path
		(v1) edge[postaction={decorate,
                   decoration={markings,mark=at position 0.53 with {\arrow{>}}}},bend right,in=170,out=10] (v2)
		(v2) edge[postaction={decorate,
                   decoration={markings,mark=at position 0.53 with {\arrow{>}}}},bend right,in=170,out=10] (v3)
        (v3) edge[postaction={decorate,
                   decoration={markings,mark=at position 0.53 with {\arrow{>}}}},bend right,in=170,out=10] (v4)
		(v4) edge[postaction={decorate,
                   decoration={markings,mark=at position 0.53 with {\arrow{>}}}},bend right,in=170,out=10] (v1)
		(v1) edge[postaction={decorate,
                   decoration={markings,mark=at position 0.53 with {\arrow{>}}}},bend right,in=170,out=10] (v3)
        (v2) edge[postaction={decorate,
                   decoration={markings,mark=at position 0.53 with {\arrow{>}}}},bend right,in=170,out=10] (v4)

        (v1) edge[postaction={decorate,
                   decoration={markings,mark=at position 0.53 with {\arrow{>}}}},bend right,in=170,out=10] (v4)
		(v4) edge[postaction={decorate,
                   decoration={markings,mark=at position 0.53 with {\arrow{>}}}},bend right,in=170,out=10] (v3)
        (v3) edge[postaction={decorate,
                   decoration={markings,mark=at position 0.53 with {\arrow{>}}}},bend right,in=170,out=10] (v2)
		(v2) edge[postaction={decorate,
                   decoration={markings,mark=at position 0.53 with {\arrow{>}}}},bend right,in=170,out=10] (v1)
		(v3) edge[postaction={decorate,
                   decoration={markings,mark=at position 0.53 with {\arrow{>}}}},bend right,in=170,out=10] (v1)
        (v4) edge[postaction={decorate,
                   decoration={markings,mark=at position 0.53 with {\arrow{>}}}},bend right,in=170,out=10] (v2)
                ;
\end{tikzpicture}\]
\end{minipage}
\caption{Undirected graphs and the associated symmetric directed graphs: path, cycle and complete graphs}
\label{ZFS:f:special}
\end{figure}

As we mentioned earlier, controllability of systems of the form \eqref{ZFS:e:sys-laplacian} has been extensively studied in the literature. In particular, minimum number of leaders that render the system \eqref{ZFS:e:sys-laplacian} controllable was investigated for some special classes of undirected graphs.
To apply our results to the special case of undirected graphs, we identify an undirected graph $H$ by a corresponding directed graph $G$ whose arc set is symmetric. As an example, three undirected graphs together with  the corresponding directed graphs are depicted in Figure~\ref{ZFS:f:special}.
For an undirected graph $H$, we denote the corresponding directed graph by $\theta(H)$. Note that, clearly, the Laplacian matrix $L$ of $H$ belongs to the qualitative class $\calQ(\theta(H))$.

In case of an undirected path graph $P_n$ with $n$ vertices, it has been shown in \cite{ARMJMMME09} that $\lmin(L)=1$. For an undirected cycle graph
$C_n$, it has been shown in \cite[Thm. 3]{SZMKCMC11} that $\lmin(L)=2$ , and any two neighbors can be chosen as leaders.
For an undirected complete graph $K_n$ with $n$ vertices, we have $\lmin(L)=n-1$, and any $n-1$ vertices can be chosen as leaders
(see \cite[Thm. 4]{SZMKCMC11}). By looking at Figure \ref{ZFS:f:special}, it is easy to verify that
$\lmin(L)$ coincides with the zero forcing number in these three cases, i.e. path, cycle, and complete graphs. Note that the set $\{1\}$ or $\{3\}$ is a minimal ZFS for the path graph in Figure \ref{ZFS:f:special}. Moreover, any two neighboring vertices constitutes a minimal zero forcing set for the cycle graph, and any three out of the four vertices forms a minimal ZFS for the complete graph
in Figure \ref{ZFS:f:special}.
Obviously, this is not limited to the depicted examples, and holds true for any undirected path, cycle, or complete graphs. Therefore, we obtain that $Z(\theta(P_n))= 1$, $Z(\theta(C_n))= 2$, and $Z(\theta(K_n))= n-1$. Then, by Corollary~\ref{ZFS:c:lmin-zfn}, we conclude that
the existing results for the minimum number of leaders rendering the system \eqref{ZFS:e:sys-laplacian} controllable, carries
over unchanged to the class of systems whose dynamics is given by \eqref{ZFS:e:sys-graph}.That is,
we have $\lmin(X)=1$ for any $X\in\calQ(\theta(P_n))$, $\lmin(X)=2$ for any $X\in\calQ(\theta(C_n))$, and $\lmin(X)=n-1$ for any $X\in\calQ(\theta(K_n))$.

It is worth mentioning that one should not conjecture based on the aforementioned special cases that $\lmin(L)$ is equal to the zero forcing number for any graph.
As a counter example, consider a 6-regular circulant graph with $10$ vertices. It follows directly from \cite[Thm. III.1]{MNMMta} that $\lmin(L)=2$,
whereas it is easy to observe that no pair of vertices results in a zero forcing set.

After the discussion of undirected graph classes for which $\lmin(L)$ has been characterized in the literature, we turn our attention to a class of directed graphs, namely directed trees (ditrees).
We use the symbol $T$ to denote a ditree to avoid possible confusion with the general case. The notions of a path, the path cover number, and a minimal path cover are required before stating the result for this case (see e.g. \cite{LH10}) for more details on these notions).

\begin{definition}
A {\em path} $P$ in $G$ is an ordered set of distinct vertices $(v_1, v_2, \ldots, v_k)$ of G such that $(v_i, v_{i+1}) \in E(G)$
for each $i=1, 2, \ldots, k-1$. The vertex $v_1$ is called {\em the initial point} of $P$ and $v_k$ is the {\em final point} of
$P$. The {\em path cover number} of $G$, denoted by $P(G)$, is the minimum number of vertex disjoint paths occurring as induced subgraphs of $G$ that cover all the
vertices of $G$; such a set of paths realizing $P(G)$ is called a {\em minimal path cover}.
\end{definition}

Now, we have the following result in case of tree structures.

\begin{proposition}\label{ZFS:p:tree}
Let $T$ be a ditree. Then, we have $\lmin(T)=P(T)$. Moreover, the initial points of the vertex disjoint paths realizing a minimal path cover form a minimal zero forcing set.
\end{proposition}

\BP
The result follows directly by applying Theorem \ref{ZFS:t:zfs-contr-directed} and Corollary \ref{ZFS:c:lmin-zfn}, together with \cite[Thm. 3.5]{LH10} and the proof provided therein.
\EP


\section{Conclusion}\label{ZFS:s:conclusion}

Controllability of systems defined on graphs has been discussed in this paper.
We have considered the problem of controllability of the network for a family of matrices carrying the structure of an underlying directed graph.
This family of matrices is called the qualitative class, and  as observed,
there is a one-to-one correspondence between the set of leaders rendering the network controllable for all matrices in the qualitative class and zero forcing sets.
We have also dealt with the case where one is interested in some subset of this qualitative class, through the notion of sufficiently rich subclasses.
To further illustrate the proposed results, special cases including path, cycle, and complete graphs are discussed.
In addition, we have shown how the proposed results of the present paper together with the existing results on the zero forcing sets lead to a minimal leader selection scheme in particular cases, such as graphs with a tree structure. Based on the results of the present paper, our knowledge about (minimal) leader selection for controllability of a network
is intimately related to the knowledge we have for zero forcing sets (number). Indeed, for each class of graphs whose zero forcing number has been known or
will be established later on, we immediately obtain the minimum number of leaders for controllability, and, in principle, a minimal leader selection scheme.

\bibliographystyle{plain}
\bibliography{ref}

\begin{thebibliography}{10}

\bibitem{ABJJ07}
A.~Banerjee and J.~Jost.
\newblock On the spectrum of the normalized graph.
\newblock {\em arXiv:0705.3772}, 2007.

\bibitem{FBWBSMFHTHLHBSPDHH10}
F.~Barioli, W.~Barrett, S.~M. Fallat, H.~T. Hall, L.~Hogben, B.~Shader,
  P.~Van~Den Driessche, and H.~Van~Der Holst.
\newblock Zero forcing parameters and minimum rank problems.
\newblock {\em Linear Algebra and its Applications}, 433(2):401--411, 2010.

\bibitem{DBSBCBVG09}
D.~Burgarth, S.~Bose, C.~Bruder, and V.~Giovannetti.
\newblock Local controllability of quantum networks.
\newblock {\em Physical Review A}, 79(6):060305(R), 2009.

\bibitem{DBDDLHSSMYta}
D.~Burgarth, D.~D'Alessandro, L.~Hogben, S.~Severini, and M.~Young.
\newblock Zero forcing, linear and quantum controllability for systems evolving
  on networks.
\newblock {\em IEEE Transactions on Automatic Control}, 58:2349 -- 2354, 2013.

\bibitem{DBVG:07}
D.~Burgarth and V.~Giovannetti.
\newblock Full control by locally induced relaxation.
\newblock {\em Physical Review Letters}, 99(10):100501(R), 2007.

\bibitem{MKCSZMC12}
M.~K. Camlibel, S.~Zhang, and M.~Cao.
\newblock Comments on `{C}ontrollability analysis of multi-agent systems using
  relaxed equitable partitions'.
\newblock {\em International Journal of Systems, Control and Communications},
  4(1/2):72--75, 2012.

\bibitem{ACMMs}
A.~Chapman and M.~Mesbahi.
\newblock Strong structural controllability of networked dynamics.
\newblock Submitted.

\bibitem{JMDCCJW:03}
J.M. Dion, C.~Commault, and J.~van~der Woude.
\newblock Generic properties and control of linear structured systems: a
  survey.
\newblock {\em Automatica}, 39(7):1125--1144, 2003.

\bibitem{MESMMCMKCAB12}
M.~Egerstedt, S.~Martini, M.~Cao, M.~K. Camlibel, and A.~Bicchi.
\newblock Interacting with networks: {H}ow does structure relate to
  controllability in single-leader, consensus network?
\newblock {\em Control Systems Magazine}, 32(4):66--73, 2012.

\bibitem{CDG10}
C.~D. Godsil.
\newblock Control by quantum dynamics on graphs.
\newblock {\em Physical Review A}, 81(5):052316--1:5, 2010.

\bibitem{LH10}
L.~Hogben.
\newblock Minimum rank problems.
\newblock {\em Linear Algebra and its Applications}, 432:1961--1974, 2010.

\bibitem{CTL74}
C.~T. Lin.
\newblock Structural controllability.
\newblock {\em IEEE Transactions on Automatic Control}, 19(3):201--208, 1974.

\bibitem{YYLJJSALB11}
Y.~Y. Liu, J.~J. Slotine, and A.~L. Barabasi.
\newblock Controllability of complex networks.
\newblock {\em Nature}, 473:167--173, 2011.

\bibitem{SMMEAB10}
S.~Martini, M.~Egerstedt, and A.~Bicchi.
\newblock Controllability analysis of multi-agent systems using relaxed
  equitable partitions.
\newblock {\em International Journal of Systems, Control and Communications},
  2(1/2/3):100--121, 2010.

\bibitem{MMME10}
M.~Mesbahi and M.~Egerstedt.
\newblock {\em Graph Theoretic Methods in Multiagent Networks}.
\newblock Princeton Series in Applied Mathematics. Princeton University Press,
  Princeton and Oxford, 2010.

\bibitem{MNMMta}
M.~Nabi-Abdolyousefi and M.~Mesbahi.
\newblock On the controllability properties of circulant networks.
\newblock {\em IEEE Transactions on Automatic Control}.
\newblock To appear.

\bibitem{GPGN12}
G.~Parlangeli and G.~Notarstefano.
\newblock On the reachability and observability of path and cycle graphs.
\newblock {\em IEEE Transactions on Automatic Control}, 57(3):743--748, 2012.

\bibitem{ARMJMMME09}
A.~Rahmani, M.~Ji, M.~Mesbahi, and M.~Egerstedt.
\newblock Controllability of multi-agent systems from a graph theoretic
  perspective.
\newblock {\em SIAM Journal on Control and Optimization}, 48(1):162--186, 2009.

\bibitem{HGT04}
H.~G. Tanner.
\newblock On the controllability of nearest neighbor interconnections.
\newblock In {\em Proc. of the 43rd IEEE conference on Decision and Control},
  pages 2467--2472, 2004.

\bibitem{MTJCDs}
M.~Trefois and J.~C. Delvenne.
\newblock Zero forcing sets, constrained matchings and minimum rank.
\newblock {\em Linear and Multilinear Algebra}.
\newblock Submitted.

\bibitem{AYYWAME12}
A.~Y. Yazicioglu, W.~Abbas, and M.~Egerstedt.
\newblock A tight lower bound on the controllability of networks with multiple
  leaders.
\newblock In {\em Proc. of the 51st IEEE conference on Decision and Control},
  pages 1978--1983, 2012.

\bibitem{SZMKCMC11}
S.~Zhang, M.~K. Camlibel, and M.~Cao.
\newblock Controllability of diffusively-coupled multi-agent systems with
  general and distance regular topologies.
\newblock In {\em Proc. of the 50th IEEE conference on Decision and Control and
  2011 European Control Conference}, pages 759--764, 2011.

\end{thebibliography}
\end{document}